\documentclass[aps,prb,twocolumn,superscriptaddress]{revtex4}
\usepackage{amsfonts}
\usepackage{amssymb}
\usepackage{amsmath}
\usepackage{graphicx}
\usepackage{epsfig}
\usepackage{color}
\usepackage{amsmath,bm}

\begin{document}

\title{Probing the superfluid-insulator phase transition by a non-Hermitian external field}
\author{X. Z. Zhang}
\affiliation{College of Physics and Materials Science, Tianjin Normal University, Tianjin
300387, China}
\author{Z. Song}
\email{songtc@nankai.edu.cn}
\affiliation{School of Physics, Nankai University, Tianjin 300071, China}

\begin{abstract}
We study the response of a thermal state of the Hubbard-like system to
either global or local non-Hermitian perturbation, which coalesces the
degenerate ground state within the $U(1)$ symmetry breaking phase. We show
that the dynamical response of the system is strongly sensitive to the
underlying quantum phase transition (QPT) from a Mott insulator to a
superfluid state. The Uhlmann fidelity in the superfluid phase decays to a
steady value determined by the order of the exceptional point (EP) within
the subspace spanned by the degenerate ground states but remains almost
unchanged in the Mott insulating phase. It demonstrates that the phase
diagram at zero temperature is preserved even though a local probing field
is applied. Specifically, two celebrated models including the Bose-Hubbard
model and the Jaynes-Cummings-Hubbard model are employed to demonstrate this
property in the finite-size system, wherein fluctuations of the boson and
polariton number are observed based on EP dynamics. This work presents an
alternative approach to probe the superfluid-insulator QPT at non-zero
temperature.
\end{abstract}

\maketitle

\section{Introduction}

In equilibrium and at zero temperature, the quantum phase transition (QPT)
serving as one of the central issues in condensed matter physics can be
usually described by a phenomenological order parameter according to the
Landau-Ginzburg theory \cite{Sachdev2011}. Therefore, a system experiences a
symmetry breaking from one phase with a nonzero order parameter to another
with a vanishing order parameter. The underlying mechanism is the degeneracy
of the ground states. Thanks to the incredible advance in quantum
simulation, especially in the context of quantum optics and atomic physics,
a wide range of condensed matter systems have been theoretically
investigated and many proposals for probing the QPT have been proposed \cite%
{Greentree2006,Hartmann2008,Koch2009}. QPTs might still be observed at
sufficiently low temperatures, where the quantum fluctuations dominate and
thermal fluctuations are not significant enough to excite the system from
its ground state. At higher temperatures, thermal fluctuations conceal the
quantum criticality. As a consequence, it leaves no residuals of the quantum
phase diagram at absolute zero temperature.

Dissipation is ubiquitous in nature and plays an essential role in quantum
systems such as inducing decoherence of quantum states. Recently, a
promising research direction is to investigate the effect of the
non-Hermiticity on the QPT \cite%
{Liu2020,Zeng2020,Guo2020,Zhang2020,Longhi2021,Zhang2021,Wu2021} and hence
discover novel quantum matters. On the other hand, much attention has been
paid to an intriguing possibility of dissipation as an efficient tool for
the preparation and manipulation of quantum states \cite%
{Mueller2012,Daley2014,Tomita2017,Zhang2020a,Nakagawa2020}. In this new
area, understanding and controlling nonequilibrium dynamics of correlated
quantum many-body systems with dissipation are an urgent issue in diverse
fields of physics, ranging from ultracold gases \cite{Sieberer2016},
Bose-Einstein condensates (BECs) placed in optical cavities \cite{Ritsch2013}%
, trapped ions \cite{Blatt2012,Bohnet2016}, exciton-polariton BEC \cite%
{Carusotto2013}, and microcavity arrays coupled with superconducting qubits
\cite{Houck2012,Fitzpatrick2017}. Given the above two fruitful topics, we
naturally ask the following questions: Can we establish a non-Hermitian
dynamic detection scheme to capture the phase of the Hermitian system and
accurately predict the phase boundary?

The QPT and the corresponding critical phenomena can be understood with the
concepts from quantum information, i.e. the quantum entanglement \cite%
{Horodecki2009,Eisert2010}, the quantum fidelity \cite%
{Zanardi2006,You2007,CamposVenuti2007}, and the Loschmidt echo (LE) \cite%
{Quan2006,Zanardi2007,Cozzini2007,Heyl2013,Abeling2016,Jafari2017,Mera2018,tang2021dynamical}%
. This provides a method for detecting QPT based on the response of the
ground state under a perturbation. The recent development of the
non-Hermitian Hamiltonian shows that it exhibits exclusive effects never
before observed in a Hermitian system \cite%
{Mostafazadeh2009,Longhi2014,Jin2018,Zhang2020a}. One of the most
interesting phenomena is the critical dynamics based on the exceptional
point (EP) \cite{Zhang2020,Zhang2021,Yang2021}. It may shed light to address
the proposed question. In this paper, we propose a scheme to detect the Mott
insulator-superfluid QPT based on the EP dynamics. In its essence, if there
can exist a non-Hermitian perturbation relating the degenerate states with
each other so as to form a Jordan block, then the order of EP can be
arbitrarily modulated according to the degeneracy of the involved states.
The EP drives the system to evolve towards the corresponding coalescent
state. Unlike the Hermitian system, the system evolution shows directional
rather than periodic oscillations even though an initial thermal state is
prepared. Based on this mechanism, we examine the response of two celebrated
Hubbard-like systems to the external critical non-Hermitian field. It
demonstrates that when the system is in the superfluid phase, the
non-Hermitian external field forces the degenerate ground state to coalesce
and thereby leads to a decay of the Loschmidt echoes. In the thermodynamic
limit, it converges to zero but stays around $1$ in the Mott insulating
phase. This dynamical property holds at a low temperature limit and is
insensitive to whether the external field is localized, and whether the
external field is isotropic. Therefore, it provides a reliable scheme for
detecting the Mott insulator-superfluid phase transition in a real physical
system.

Our paper is structured as follows: In Sec. II, we give the fundamental
mechanism of the proposed non-Hermitian detecting scheme through a simple
two-level system. In Sec. III and IV, we apply the proposal to examine the
Mott insulator-superfluid QPT in two celebrated correlated many-body
systems, namely, the Jaynes-Cummings-Hubbard (JCH) model and Bose-Hubbard
(BH) model. We conclude and discuss our results in Sec. V.

\section{Insight into the Non-Hermitian detection}

\begin{figure}[tbp]
\centering
\includegraphics[width=0.4\textwidth]{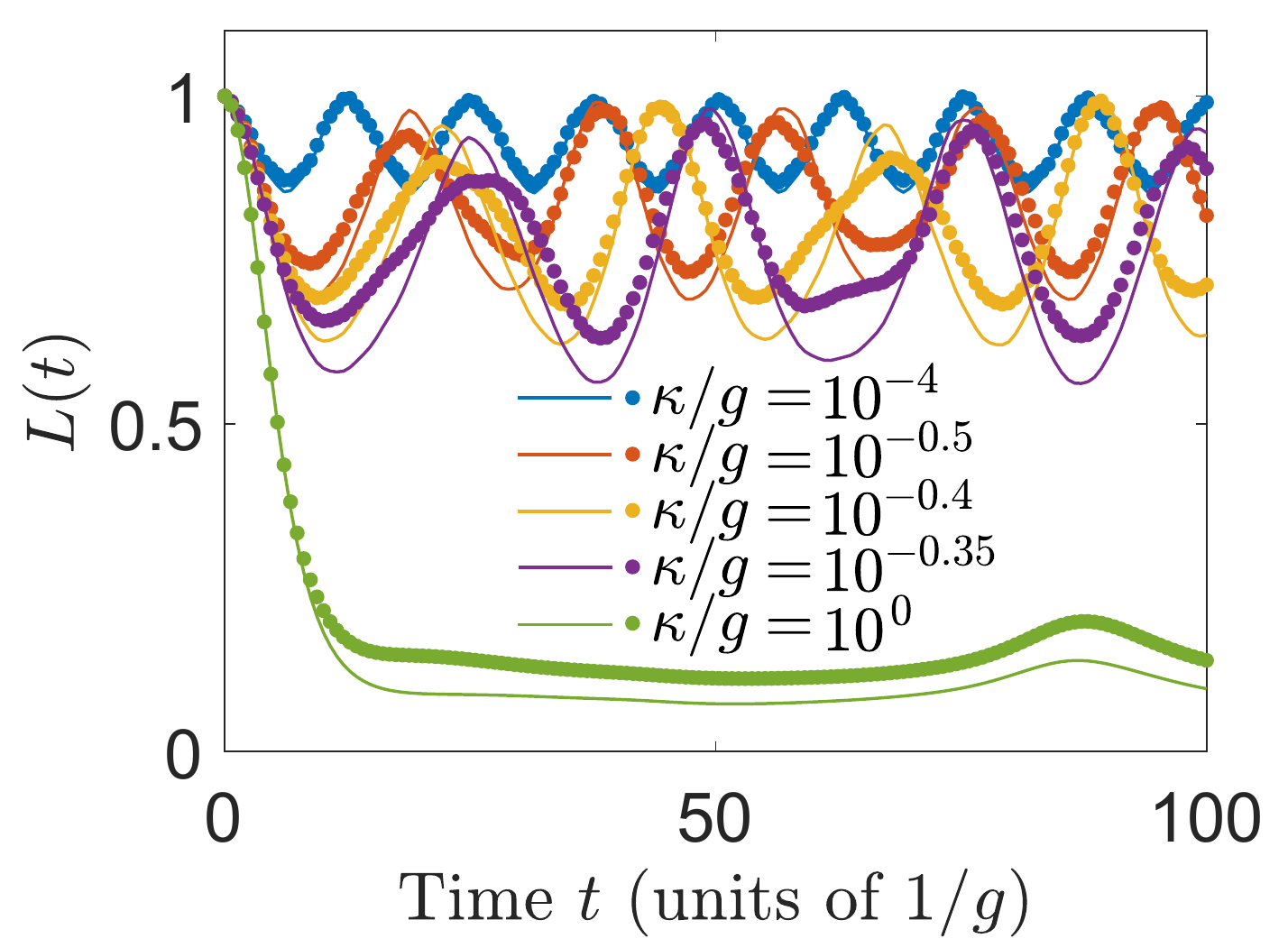}
\caption{Time evolution of LEs for different $\protect\kappa $ with a given
chemical potential $\protect\mu $. The lines and dots denotes the LEs for $%
\protect\beta =5$ and $\protect\beta =10$, respectively. The system consists
of $2$ cavities, the photon number of each cavity is truncated at a finite
value $n=10$. The other system parameters are $g=1$, $\protect\omega _{c}=5g$%
, $\Delta =0$, $\protect\lambda =0.1g$ and $\protect\mu =-6.5g$. The
profiles of the LEs in the two regions are distinct, independent of the
temperature of the initial thermal states, and converge to $1.0$ and $0.2$,
respectively.}
\label{fig1}
\end{figure}
We first demonstrate the underlying mechanism of the considered proposal
through a simple $2\times 2$ matrix. The starting point is a Hermitian
two-level system with the eigenenergies being $E_{1}$ and $E_{2}$. In the
energy representation $\left\{ \left\vert \psi _{1}\right\rangle \text{, }%
\left\vert \psi _{2}\right\rangle \right\} $, the matrix form can be given
as
\begin{equation}
H_{0}=\left(
\begin{array}{cc}
E_{1} & 0 \\
0 & E_{2}%
\end{array}%
\right) ,
\end{equation}%
where the two energy levels are arranged in ascending order such
that $E_{1}<E_{2}$. We focus on the dynamics of a initial thermal state with
density matrix $\rho \left( 0\right) =e^{-\beta H_{0}}/\mathrm{Tr}(e^{-\beta
H_{0}})$ at temperature $T=1/\beta $ with $\hbar =1$. Evidently, when the
two energy levels are near degenerate and low temperature limit is assumed,
the density matrix $\rho \left( 0\right) $ is reduced to $\rho _{\mathrm{I}%
}\left( 0\right) =I/2$ where $I$ is identity matrix. On the contrary, if
there exists a gap $\delta =E_{2}-E_{1}$ between the involved two energy
levels, then the initial density matrix of the system is reduced to $\rho _{%
\mathrm{II}}\left( 0\right) =(I+\sigma _{z})/2$. Note in passing that the
first type of initial state is a maximally mixed state demonstrated by $%
\mathrm{Tr}[\rho _{\mathrm{I}}^{2}\left( 0\right) ]=1/2$ and the second type
of initial thermal state is a pure state characterized by $\mathrm{Tr}(\rho
_{\mathrm{II}}\left( 0\right) )^{2}=1$. The interplay between $\delta$ and $%
\beta$ determines the constituents of each eigenstate in the mixed state.
For instance, the larger $\beta$ is required to involve the information of
excited state in the initial thermal state when $\delta$ is large. These
evidences play the key role to understand the quench dynamics. After a
non-Hermitian quench, the post-quench Hamiltonian can be given as $%
H=H_{0}+H^{\prime }$, wherein $H^{\prime }=\lambda \left\vert \psi
_{1}\right\rangle \left\langle \psi _{2}\right\vert $. The corresponding
matrix form is%
\begin{equation}
H=\left(
\begin{array}{cc}
E_{1} & \lambda \\
0 & E_{2}%
\end{array}%
\right),
\end{equation}%
where $\lambda $ is a real number and denotes the non-Hermitian coupling
between two such energies. When the two energies of the pre-quench
Hamiltonian are degenerate $E_{1}=E_{2}$, the post-quench Hamiltonian $H$ is
in a Jordan block form such that the two eigenstates $\left\vert \varphi
_{1}\right\rangle $\ and $\left\vert \varphi _{2}\right\rangle $ of $H$
coalesce. For any given pure initial state, the quenched Hamiltonian drive
it to the coalescent state (see Appendix for details). However, when $\delta
\gg 1$, the quenched Hamiltonian $H$ shares the same spectrum with $H_{0}$;
the eignestate $\left\vert \varphi _{1}\right\rangle $ is unchanged even
though a non-zero perturbation $\lambda $ presents; the eigenstate $%
\left\vert \varphi _{2}\right\rangle $ of $H$ is in a superposition of $%
\left\vert \psi _{1}\right\rangle $ and $\left\vert \psi _{2}\right\rangle $%
, that is $\left\vert \varphi _{2}\right\rangle =(\lambda /\delta
)\left\vert \varphi _{1}\right\rangle +\left\vert \varphi _{2}\right\rangle $%
. It is conceivable that the dynamics of the two initial thermal states will
exhibit distinct behaviors. To give such differences, we first investigate
the time evolution of the density matrix $\rho \left( t\right) $. It should
obey the following equation
\begin{equation}
i\frac{\partial \rho \left( t\right) }{\partial t}=H\rho \left( t\right)
-\rho \left( t\right) H^{\dagger },
\end{equation}%
which admits the formal solution%
\begin{equation}
\rho \left( t\right) =e^{-iHt}\rho \left( 0\right) e^{iH^{\dagger }t}.
\end{equation}%
Due to the non-Hermiticity nature, the time evolution of the density matrix
is no longer unitary. Hence, in the subsequent analysis, we normalize $\rho
\left( t\right) $ by taking
\begin{equation}
\rho \left( t\right) =e^{-iHt}\rho \left( 0\right) e^{iH^{\dagger }t}/%
\mathrm{Tr}(e^{-iHt}\rho \left( 0\right) e^{iH^{\dagger }t}).
\end{equation}%
The degree of distinguishability between the initial state $\rho \left(
0\right) $ and $\rho \left( t\right) $ can be identified by the so-called
Uhlmann fidelity \cite{Uhlmann1976,Jacobson2011}
\begin{equation}
L\left( t\right) =[\mathrm{Tr}(\rho ^{1/2}\left( 0\right) \rho \left(
t\right) \rho ^{1/2}\left( 0\right) )^{1/2}]^{2},  \label{Le}
\end{equation}%
also known as Loschmidt echo (LE).
\begin{figure*}[tbp]
\centering
\includegraphics[width=0.95\textwidth]{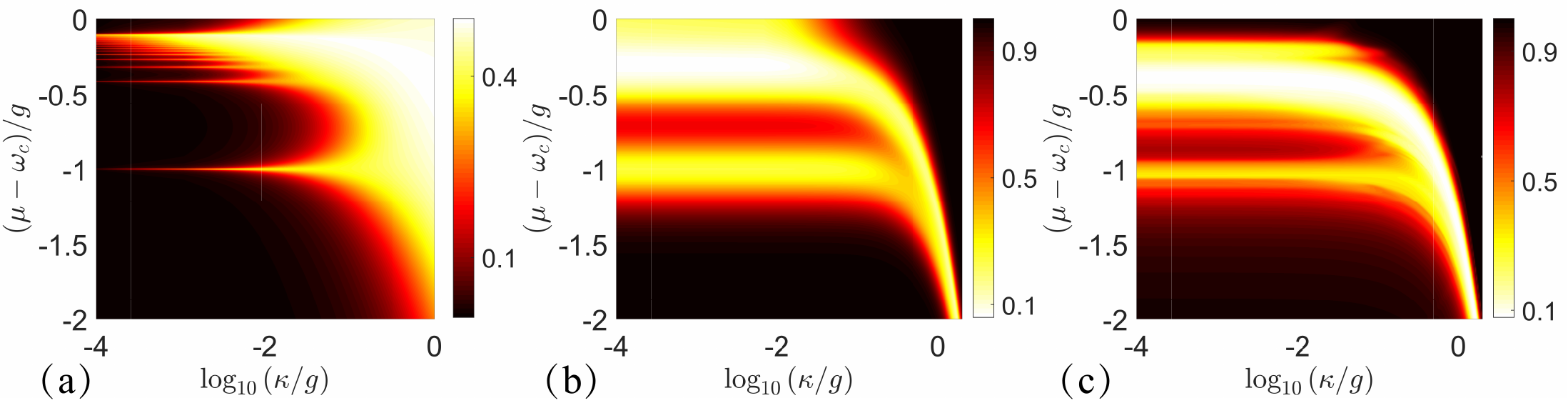}
\caption{Comparison of phase diagrams obtained by the mean-field
approximation, purity, and average LEs. Here we set $T=200$. Other
parameters: $g=1$, $\protect\lambda =0.1$, and $\protect\beta =10$. The
behaviors of $\overline{L}$ and purity accord with each other and
demonstrate the phase boundary even in the finite-size system. It also
presents a clear manifestation of the Mott insulator to superfluid quantum
phase transition at non-zero temperature. Note that in the upper right part
of Figs. (b)-(c), the system possesses a gap due to the finite size effect,
which can be expected to vanish as $N$ increases to infinite. Note that the
colorbar of (a) is reversed as opposed to that of (b) and (c). }
\label{fig2}
\end{figure*}
For the first type of initial state $\rho _{\mathrm{I}}\left( 0\right) $,
the straightforward algebra shows that%
\begin{equation}
\rho _{\mathrm{I}}\left( t\right) =\frac{1}{\Omega _{\mathrm{I}}\left(
t\right) }\left(
\begin{array}{cc}
t^{2}\lambda ^{2}+1 & -it\lambda \\
it\lambda & 1%
\end{array}%
\right) ,
\end{equation}%
where $\Omega _{\mathrm{I}}\left( t\right) =\lambda ^{2}t^{2}+2$.
Substituting $\rho _{\mathrm{I}}\left( t\right) $ into Eq. (\ref{Le}), one
can immediately obtain
\begin{eqnarray}
L\left( t\right) &=&[\frac{1}{2\Omega _{\mathrm{I}}^{1/2}\left( t\right) }(%
\sqrt{\Omega _{\mathrm{I}}\left( t\right) +\lambda t[\Omega _{\mathrm{I}%
}\left( t\right) +2]^{1/2}}  \notag \\
&&+\sqrt{\Omega _{\mathrm{I}}\left( t\right) -\lambda t[\Omega _{\mathrm{I}%
}\left( t\right) +2]^{1/2}})]^{2}.
\end{eqnarray}%
Our primary interest here is the steady-value of LE $L\left( t\right) $ ($%
t\rightarrow \infty $) after a sufficient long period, which can be given by
setting $\lambda t_{f}\gg 1$. Within this condition, $\Omega _{\mathrm{I}%
}\left( t_{f}\right) \approx \lambda ^{2}t_{f}^{2}$, and hence $L\left(
t_{f}\right) \approx 1/2$. The physical picture is clear: The initial mixed
state $\rho _{\mathrm{I}}\left( 0\right) $ contains components of two
parities. When the non-Hermitian coupling $\lambda $ is switched on, the
post-quench non-Hermitian Hamiltonian $H$ contains only one coalescent state
$\left\vert \varphi _{\mathrm{c}}\right\rangle =\left\vert \psi
_{1}\right\rangle $. Therefore, all the possible initial states will be
driven towards this coalescent state. This indicates that the component with
a certain parity ($\left\vert \psi _{1}\right\rangle $) of the thermal state
$\rho _{\mathrm{I}}\left( 0\right) $ is dominant since the EP dynamics. From
this perspective, $\rho _{\mathrm{I}}\left( t\right) $ loses half of the
information regarding the $\left\vert \psi _{2}\right\rangle $, which
results in $L\left( t_{f}\right) \approx 1/2$. These features do not occur
when the Hermitian field $H^{\prime }$ is applied since $L\left( t\right) $
is always $1$ as time $t$ goes by. These conclusions still hold for the
dynamical detection scheme of $N$-fold degenerate system. In that setup, the
non-Hermitian detection filed is given as $H^{\prime }=\lambda
\sum_{j=1}^{N-1}\left\vert \psi _{j}\right\rangle \left\langle \psi
_{j+1}\right\vert $ and post-quench Hamiltonian $H$ possesses only one
eigenvalue whose geometric multiplicity being $1$. Hence, the high-order EP
point up to $N$-level coalescence is created. Any given arbitrary initial
state will evolve towards the coalescent state $\left\vert \psi
_{1}\right\rangle $ after sufficiently long time. At this time, $\rho _{%
\mathrm{I}}\left( t\right) $ tends to $\left\vert \psi _{1}\right\rangle
\left\langle \psi _{1}\right\vert $ leading to $L\left( t_{f}\right) \approx
1/N$. It also demonstrates that the order of EP determines the steady-value
of LE $L\left( t\right) $.

For the second type of initial state $\rho _{\mathrm{II}}\left( 0\right) $,
the time evolution of the density matrix can be readily obtained as $\rho _{%
\mathrm{II}}\left( t\right) =\rho _{\mathrm{II}}\left( 0\right) =(I+\sigma
_{z})/2$. The LE $L\left( t\right) $ can be given directly as $L\left(
t\right) \approx L\left( 0\right) =1$. This denotes that a non-Hermitian
detection does not substantially affect the dynamics due to the presence of
gap $\Delta $. It is worth pointing out that if the Hermitian detection
field $H^{\prime }=\lambda \left\vert \psi _{1}\right\rangle \left\langle
\psi _{2}\right\vert +$ \textrm{H.c.} is turned on, then LE $L\left(
t\right) =1-\lambda ^{2}\sin ^{2}\left( \omega t\right) /\omega ^{2}$
wherein $\omega =\sqrt{\Delta ^{2}+\lambda ^{2}}$. Evidently, it is a
periodic function. In the weak coupling limit $\lambda \ll 1$, $L\left(
t_{f}\right) $ stays near $1$, which is similar to that of the non-Hermitian
detection scheme. In the following, we will demonstrate that the considered
scheme can be applied to examine the QPTs, which are usually associated with
the spontaneous symmetry breaking of the system. When the system enters from
one phase to another, the system energy will undergo a transition from gap
to gapless wherein the phase transition point corresponds to the gap closing
point. Hence, the system will exhibit distinct dynamic behaviors when it is
in the different phases. Such difference can be detected by the current
proposed scheme.

\section{Jaynes-Cummings-Hubbard model}

\begin{figure*}[tbp]
\centering
\includegraphics[width=0.75\textwidth]{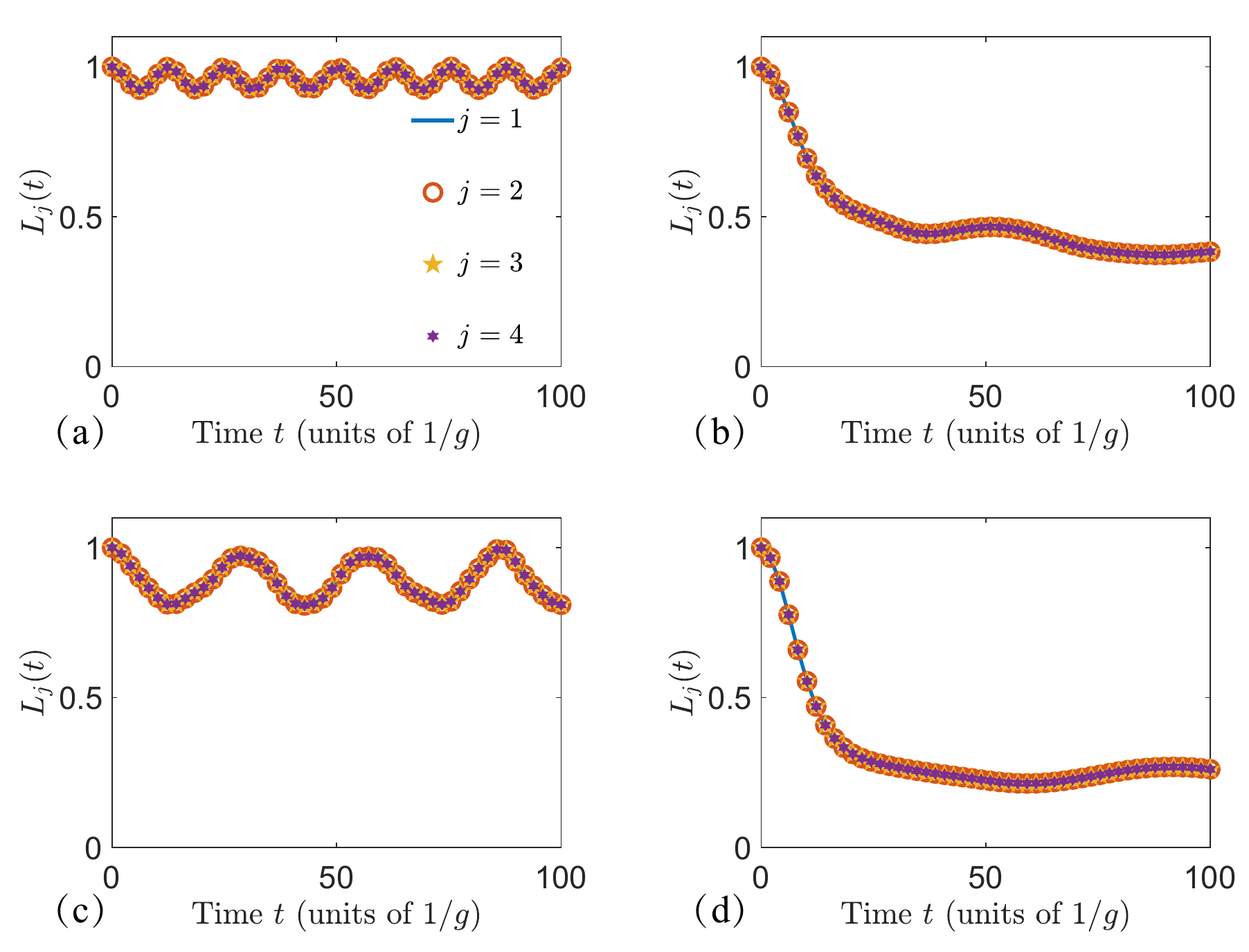}
\caption{Numerical simulation for LEs under the postquench Hamiltonian (%
\protect\ref{local_post_quench}) for different $i$ values. (a) and (c) LEs
in the Mott insulating phase when $\protect\kappa /g=10^{-4}$ and $\protect%
\kappa /g=10^{-0.08}$. (b) and (d) LEs in the superfluid phase $\protect%
\kappa /g=10^{-0.3}$ and $\protect\kappa /g=10^{0}$. The system is composed
of four cavities and the other system parameters are $\protect\mu =-6.5g$,
and $\protect\beta =10$. The LE decays rapidly to $0.25$, whereas it remains
at one in the Mott insulating phase. This evidence manifests that a local
dissipation can lead to a significantly change of $\protect\rho \left(
t\right) $ thereby serves as a dynamical signature to identify different
phases of matter.}
\label{fig3}
\end{figure*}
The first celebrated QPT model is the JCH model that has emerged as a
fundamental model at the interface of quantum optics and condensed matter
physics \cite%
{Greentree2006,Buluta2009,Rabl2011,Polkovnikov2011,Chang2014,Georgescu2014,Lodahl2015}%
. It describes strongly correlated photons in a coupled qubit-cavity array
and predicts a superfluid-Mott insulator transition of polaritons. The
corresponding Hamiltonian reads
\begin{equation}
H_{0}=\sum_{i}H_{i}^{\mathrm{JC}}+\sum_{<i,j>}\kappa _{ij}(a_{i}^{\dagger
}a_{j}+\text{\textrm{H.c.}})-\sum_{i}\mu _{i}N_{i},
\end{equation}%
with
\begin{eqnarray}
H_{i}^{\mathrm{JC}} &=&\omega _{c}a_{i}^{\dagger }a_{i}+\omega _{a}\sigma
_{i}^{+}\sigma _{i}^{-}+g\left( a_{i}\sigma _{i}^{+}+a_{i}^{\dagger }\sigma
_{i}^{-}\right) , \\
N_{i} &=&\sigma _{i}^{+}\sigma _{i}^{-}+a_{i}^{\dagger }a_{i},
\end{eqnarray}%
where $\sigma _{i}^{+}=|e_{i}\rangle \langle g_{i}|$ and $\sigma
_{i}^{-}=|g_{i}\rangle \langle e_{i}|$ ($a_{i}^{\dagger }$, $a_{i}$)
correspond to the atomic (photonic) raising and lowering operators,
respectively. $|g_{i}\rangle $, $|e_{i}\rangle $ are the ground and excited
states of the two-level system. The transition energy of the atomic system
is $\omega _{a}$, the cavity resonance is $\omega _{c}$ and the cavity
mediated atom-photon coupling is $g$, which is assumed to be real for our
purposes. The whole system is given by a combination of the Jaynes-Cummings
Hamiltonian $H_{i}^{\mathrm{JC}}$ with photon hopping between cavities $%
\kappa _{ij}$ ($\langle i,j\rangle $ represents nearest-neighbour pairs) and
the chemical potential term $\mu _{i}$. Here $N=\sum_{i}N_{i}$ is the total
number of atomic and photonic excitations, which is a conserved quantity,
i.e., $\left[ N,\text{ }H\right] =0$. This is also called $U(1)$ symmetry,
conserves the number of polaritons.

For simplicity, we assume that the homogeneous inter-cavity hopping $\kappa
_{ij}=\kappa \delta _{i,j+1}$ occurs for nearest neighbours and zero
disorder $\mu _{i}=\mu $ for all sites. Because of the photonic repulsion
arising from $g$, the system supports two phases, that is, the Mott
insulating and superfluid phases. Such phases can be determined by employing
a mean-field approximation. Although the mean-field theory as an
approximation theory is not particularly accurate, it can give the basic
property of the ground state of two phases. To capture such property, we
first give the dressed states $|\pm ,n\rangle $ of $H^{\mathrm{JC}}$ (the
subscript $i$ is omitted), where $n$ is the number of excitations in the
cavity. The concrete forms of such states can be given as
\begin{equation}
|\pm ,n\rangle =\frac{g\sqrt{n}|g,n\rangle +[-\Delta /2\pm \chi \left(
n\right) ]|e,n-1\rangle }{\sqrt{2\chi ^{2}\left( n\right) \pm \chi \left(
n\right) \Delta }}\text{ }\forall n\geqslant 1\text{,}
\end{equation}%
and the corresponding eigenenergies are
\begin{equation}
E_{\pm ,n}=n\omega _{c}\pm \chi \left( n\right) -\Delta /2,
\end{equation}%
where detuning $\Delta =\omega _{c}-\omega _{a}$ and $\chi \left( n\right) =%
\sqrt{ng^{2}+\Delta ^{2}/4}$. The ground state for the dressed state system
is defined as $|g,0\rangle $ with eigenenergy $E_{\mathrm{g}}=0$. Taking the
decoupling approximation $a_{i}^{\dagger }a_{j}=\psi ^{\ast }a_{j}+\psi
a_{i}^{\dagger }-|\psi |^{2}$ with $\psi =\langle a_{i}\rangle $, we can
demonstrate that when the system is in the Mott insulating phase $\psi =0$,
the ground state of the system has a fixed number of polaritonic excitation
on each site, which is determined by system parameters. There must be a gap
between the ground state and the first excited state of the system. As a
comparison, the system is in the superfluid phase when $\psi \neq 0$. At
this time, the system is gapless and the ground state at each site
corresponds to a coherent state of excitations over $|-,n\rangle $ branch.
Note that the condition of $E_{-,n}<E_{+,n}$ is assumed. These properties
allow us to dynamically identify two such phases by employing the
non-Hermitian probing field that can be given as $H^{\prime }=\lambda
\sum_{i}a_{i}$ in this scenario. After a quench, one can expect
that $L\left( t\right) $ of the initial thermal state with a low-temperature
limit will not decay due to the protection of the gap. On the contrary, when
the system is tuned to the superfluid phase, the ground state is forced to
be degenerate to break the symmetry. The degenerate ground states possessing
the different excitation numbers can be related to each other through $%
H^{\prime }$ such that a Jordan block form appears. The degeneracy of the
ground state determines the order of the EP of $H$. In the thermodynamic
limit, the steady-value of $L\left( t\right) $ quickly approaches $0$
according to the EP$\ $dynamics of $\rho _{\mathrm{I}}\left( t\right) $ in
the aforementioned section. In the finite-size system, the change of the
ground state symmetry accords with that predicted by the mean-field theory,
but the exact phase boundary cannot be determined by that approximation. As
a benchmark, the purity $\mathrm{Tr}\left[ \rho ^{2}\left( 0\right) \right] $
is employed to identify whether the ground state is degenerate in the
low-temperature limit. Evidently, $\mathrm{Tr}\left[ \rho ^{2}\left(
0\right) \right] =1$ when the ground state is not degenerate. On the other
hand, the presence of the degenerate ground states makes the purity tend to $%
1/N_{c}$, with $N_{c}$ denoting the degeneracy. Note that there can exist a
gap in the finite-size system even though the system is in the deep
superfluid regime ($t\gg \beta $) characterized by the constant number of
correlation function $\langle a_{i}^{\dagger }a_{j}\rangle $. However, it
will vanish as the system dimension increases. This property does not affect
the validity of the current proposed non-Hermitian scheme to detect the QPT
boundary at which the excitation spectrum is gapless.

To verify the above conclusion, we perform the numerical simulations for $%
L\left( t\right) $ of the initial state $\rho \left( 0\right) $ at different
phases in the finite system. In Fig. \ref{fig1}, the non-Hermitian quenched
Hamiltonian drives the system exhibit two distinct behaviors of $L\left(
t\right) $: In the Mott insulating phase characterized by a fixed number of
excitations per site with no fluctuations, $L\left( t\right) $ will stay at $%
1$ as time $t$ goes on; in a superfluid phase, $L\left( t\right) $ tends
towards a steady value depending on the purity of the initial thermal state.
These results agree with our prediction and demonstrate that LEs are
insensitive to temperature and tend towards different values in different
phases.
\begin{figure}[tbp]
\centering
\includegraphics[width=0.4\textwidth]{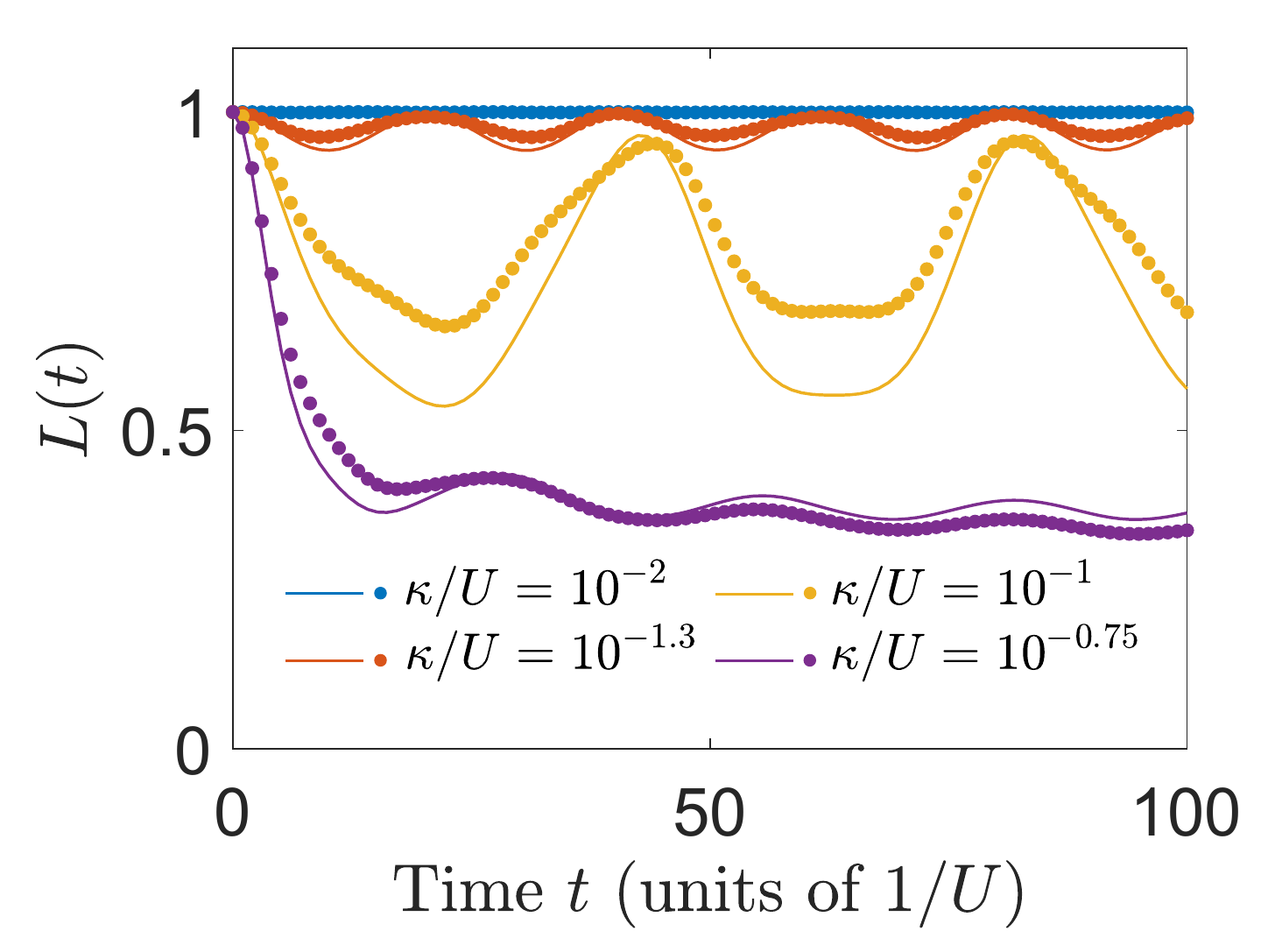}
\caption{Time evolution of LEs for different ratio of $\protect\kappa /U$.
The lines and dots denote the LEs for $\protect\beta =5$ and $\protect\beta %
=10$, respectively. The system consists of $4$ sites with photon numbers
truncated at a finite value $n=15$. The chemical potential is assumed to be $%
\protect\mu/U =0.5$ such that each site's occupation number of the ground
state is $1$ within the Mott insulating phase. The strength of the
non-Hermitian perturbation field is set to be $\protect\lambda /U=0.1$ to
induce the Jordan block form in the superfluid phase. Again, the LEs exhibit
different dynamical behaviors in two such phases: $L\left( t\right) $ tends
towards a steady value $0.2$ in the Mott insulating phase whereas it remains
$1$ in the superfluid phase. The profiles of LEs is independent of the
temperature of the initial thermal states.}
\label{fig4}
\end{figure}
To compare with the phase diagram obtained by the mean-field theory and the
purity, we introduce an average LE in the time interval $[0,T]$ that is
defined as
\begin{equation}
\overline{L}=\frac{1}{T}\int_{0}^{T}L\left( t\right) \mathrm{d}t,
\end{equation}%
where $T\gg 1$. Average LE as a function of parameters $\kappa $ and $\mu $
values with given $\Delta=0 $ is plotted in Fig. \ref{fig2}(c). Comparing to
the order parameter $\psi$ obtained by mean-field approximation in the
thermodynamic limit, it indicates that the average LE can be used to
identify the quantum phase diagram at nonzero temperatures even in small
size systems.

Now we turn to examine how does the local external field can affect the $%
L\left( t\right) $. Consider the post-quench Hamiltonian with the form%
\begin{equation}
H=H_{0}+\lambda a_{i},  \label{local_post_quench}
\end{equation}%
where $\lambda a_{i}$ is the component of operator $\lambda
\sum_{i}a_{i}$. In this case, the LE is denoted by $L_{j}\left(
t\right) $. In the superfluid phase, a local external field can indeed make
the degenerate ground states coalesce, thereby the long-term behavior of $%
L_{j}\left( t\right) $ is expected to be similar to that of the post-quench
Hamiltonian in the presence of the global non-Hermitian field. We perform
the numerical simulation in Fig. \ref{fig3}. We can see that $L_{j}\left(
t\right) $ decay to a steady value in the superfluid phase, but remain $1$
in the Mott insulating phase after a sufficiently long time. This accords
with our prediction. This evidence manifests that a local dissipation
affects qualitatively the dynamics of the initial state through EP and hence
provides a new mechanism to probe the QPT from Mott insulator to superfluid.

\begin{figure*}[tbp]
\centering
\includegraphics[width=0.8\textwidth]{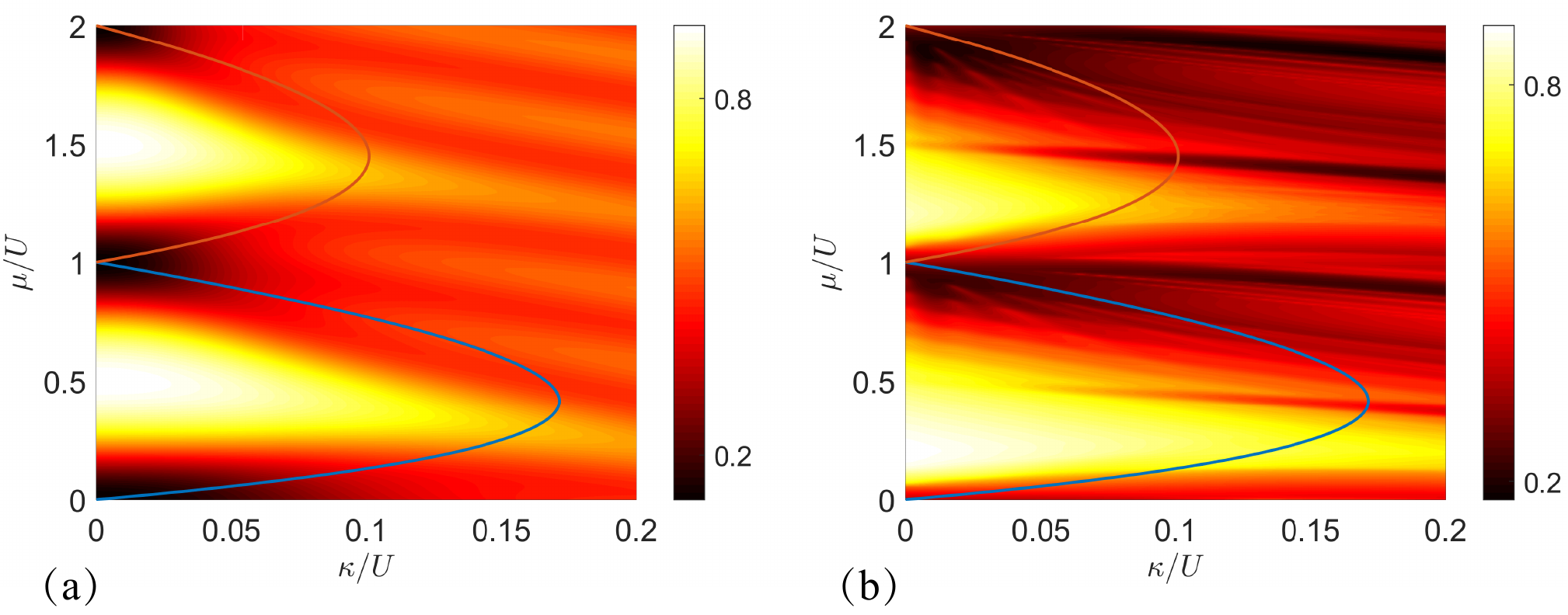}
\caption{Contour plot of average LEs $\overline{L}$ and $\overline{L}_{i}$
in (a) and (b), respectively. The Mott-lobes denoted by the red and blue
solid lines is obtained by the mean-field approximation. Here the local
dissipation is applied to site 1 and $T=200$ is set to evaluate the average
LEs. Other system parameters are (a) $\protect\lambda /U=0.1$, and $\protect%
\beta =5$; (b) $\protect\lambda /U=0.2$, and $\protect\beta =5$; It
demonstrates that the response of the thermal state to a nonlocal and local
non-Hermitian perturbation field is the same so that two such fields can be
served as signature to witness the QPT in a real experiment.}
\label{fig5}
\end{figure*}

\section{Bose-Hubbard model}

The second celebrated model delineating a Mott-insulator-superfluid
transition is the BH model. The corresponding Hamiltonian is
\begin{equation}
H_{0}=-\sum_{<i,j>}\kappa _{ij}(b_{i}^{\dagger }b_{j}+\text{\textrm{H.c.}})+%
\frac{U}{2}\sum_{i}n_{i}\left( n_{i}-1\right) -\mu \sum_{i}n_{i}.
\end{equation}%
Here $b_{i}^{\dagger }$ and $b_{i}$ are bosonic creation and annihilation
operators such that $n_{i}=b_{i}^{\dagger }b_{i}$ gives the number of
particles on-site $i$. $\kappa _{ij}$, $\mu $, and $U$ are tunable
parameters of the BH model, corresponding to the tunneling, chemical
potential, and interaction strength, respectively. The system also respects
the $U(1)$ symmetry, that is $[\sum_{i}n_{i},$ $H]=0$, which conserves the
number of bosons instead of polaritons. The BH model is closely related to
the Hubbard model which originated in solid-state physics as an approximate
description of superconducting systems and the motion of electrons between
the atoms of a crystalline solid. In the experiment of ultra-cold atom
loaded into the optical lattice, the considered model can be explored from a
superfluid to Mott insulating phases \cite{Sengupta2005,Capello2007,Bakr2010}
by addressing the laser field and manipulating Feshbach resonance \cite%
{Ohashi2002,Kevrekidis2003,Theis2004}. This model can also be used to
describe physical systems such as bosonic atoms in an optical lattice, as
well as certain magnetic insulators \cite{Jaksch2005,Giamarchi2008,Zapf2014}.

Again, the system exhibits two different phases of matter by tuning the
ratio $t/U$. The Mott insulating phase is essentially a product of
single-site states of bosons where there is a finite energy gap opposing the
addition of a boson. The excitation spectrum of the superfluid is gapless in
the sense that the sum of the energy cost needed to add and to remove one
particle from the system is zero. The superfluid phase shows boson number
fluctuations instead of polariton number fluctuations in the JCH model.
According to the value of $\psi $, the mean-field phase boundary of the BH\
model is shown in Fig. \ref{fig5}. With the same procedure, we consider the
quench dynamics of the initial therm state $\rho \left( 0\right) $. The
non-Hermitian applied field is $H^{\prime }=\lambda \sum_{i}b_{i}$%
. After a quench, we first evaluate the performance of $L\left( t\right) $
in such two different phases of matter. The evolved density matrix $\rho
\left( t\right) $ is the same as that in the JCH model, which can be shown
in Fig. \ref{fig4}. In addition, we numerically compute the $\overline{L}%
\left( t\right) $ and $\overline{L}_{i}\left( t\right) $ in the finite-size
system. Fig. \ref{fig5} shows that the Mott-lobes can be determined by $%
\overline{L}$ indicating that the phase diagram can be preserved in the
finite-size system. Note in passing that a local non-Hermitian quench field $%
H^{\prime }=\lambda b_{i}^{\dagger }$ can also dynamically identify two such
phases which can be shown by comparing Fig. \ref{fig5} (a) and (b). It paves
the way to understanding the spontaneous symmetry breaking of matter at
nonzero temperatures.

\section{Conclusion}

In conclusion, we have witnessed the Mott-insulator to superfluid phase
transition from zero to non-zero temperatures. The gapless excitation
spectrum, which serves as the signature of the $U\left( 1\right) $ symmetry
breaking, is crucial to achieving the conclusion. Such nonzero-temperature
QPT can be probed through an inhomogeneous non-Hermitian external field. The
evolved state with specific direction arising from the EP dynamics amplifies
the difference between two phases of matter, which has no counterpart in
Hermitian regime and allows distinct responses in two such phases. We expect
that the scheme proposed in this paper can be exploited to uncover as yet
unexplored Hubbard-like models in a variety of physical systems.

\acknowledgments We acknowledge the support of the National Natural Science
Foundation of China (Grants No. 11975166, and No. 11874225).

\section{Appendix}

Considering that the eigenstates of the two-level system are degenerate. In
the basis of $\left\{ \left\vert \psi _{1}\right\rangle \text{, }\left\vert
\psi _{2}\right\rangle \right\} $, the post-quenched Hamiltonian is
\begin{equation}
H=\left(
\begin{array}{cc}
E & \lambda \\
0 & E%
\end{array}%
\right) .
\end{equation}%
where $E=E_{1}=E_{2}$ is supposed. It has a Jordan block structure such that
the degenerates become coalesce with a coalescent state $\left\vert \varphi
_{\mathrm{c}}\right\rangle =(%
\begin{array}{cc}
1 & 0%
\end{array}%
)^{\mathrm{T}}$. For an arbitrary initial state$\left\vert \Phi \left(
0\right) \right\rangle =a\left\vert \psi _{1}\right\rangle +b\left\vert \psi
_{2}\right\rangle $,\ its time evolution can be determined by the propagator
$U\left( t\right) $ that has an explicit form
\begin{equation}
U\left( t\right) =e^{-iEt}\left[ I_{2}-i\lambda t\left(
\begin{array}{cc}
0 & 1 \\
0 & 0%
\end{array}%
\right) \right] .
\end{equation}%
Hence, the evolved state by neglecting the overall phase can be given as $%
\left\vert \Phi \left( t\right) \right\rangle =\left( a-ibt\lambda \right)
\left\vert \psi _{1}\right\rangle +b\left\vert \psi _{2}\right\rangle $.
After a sufficiently long time, the probability in $\left\vert \psi
_{1}\right\rangle $ overwhelms that in $\left\vert \psi _{2}\right\rangle $
ensuring the final evolved state be the coalescent state $\left\vert \varphi
_{\mathrm{c}}\right\rangle $.


\end{document}